\documentclass{svjour3}                     
\smartqed  
\usepackage{graphicx}
\journalname{Few-Body Systems (FB20)}
\begin{document}

\title{
Microscopic Optical Potentials for Helium-6 Scattering off
Protons
\thanks{Supported in part by the U.S. Department of Energy under
contract No. DE-FG02-93ER40756 with Ohio University and
under contract No. DE-SC0004084 (TORUS Collaboration).}
\thanks{Presented at the 20th International IUPAP Conference on Few-Body Problems in Physics, 20 - 25 August, 2012, Fukuoka, Japan}
}

\author{Ch. Elster         \and
        A. Orazbayev      \and
        S.P. Weppner
}

\institute{Ch. Elster \at
              Dept. of Physics and Astronomy,
              Ohio University, Athens, OH, 45701, USA \\
              \email{elster@ohio.edu}           
           \and
           A. Orazbayev \at
              Dept. of Physics and Astronomy,
              Ohio University, Athens, OH, 45701, USA \\
              \email{ao379408@ohio.edu} 
            \and
           S.P. Weppner \at
           Natural Sciences, Eckerd College, St. Petersburg, FL 33711, USA
           \email{weppnesp@eckerd.edu}
}

\date{Received: date / Accepted: date}

\maketitle

\begin{abstract}
The differential cross section and the analyzing power are calculated
for elastic scattering of $^6$He from a proton target using a
microscopic folding optical potential, in which the $^6$He nucleus
is described in terms of a $^4$He-core with two additional neutrons in
the valence p-shell.  In contrast to previous work of that nature, all
contributions from the interaction of the valence neutrons with the
target protons are taken into account.
\end{abstract}

\section{Introduction}
\label{intro}
Recently the Helium isotopes have extensively studied, both
experimentally and theoretically. Specifically,  elastic scattering of
$^6$He off a polarized proton target has been measured
for the first time at an energy of 71~MeV/nucleon~\cite{Sakaguchi:2011rp}.
The experiment finds that the analyzing power becomes negative around
50$^o$, which is not predicted by simple folding models for the optical
potential~\cite{Weppner:2000fi,Gupta:2000bu}, but which
nevertheless describe the
differential cross section at this energy reasonably well.
This apparent `$A_y$ problem' leads to the conclusion that 
folding models which are adequate in describing p-A scattering from stable,
closed shell nuclei need to be revisited and extended when applying them to 
unstable as well as open-shell nuclei.

\section{First Order Optical Potential including Valence Neutrons}
\label{sec:1}

\begin{figure}
\begin{center}
\includegraphics[width=55mm]{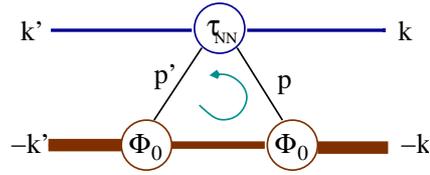}
\caption{Diagram for the optical potential matrix element in the
single-scattering approximation.}
\end{center}
\label{fig1} 
\end{figure}

The theoretical approach to elastic scattering of a nucleon from a nucleus
pioneered by Watson~\cite{Watson} has been applied very successfully over the
years to a wide range of closed shell nuclei. Here a spectator expansion is
constructed within a multiple scattering theory predicated upon the idea that
the two-body interactions between the projectile and the nucleons in the target
dominate the reaction. Generally, only the first order in this expansion is
considered, leading to a Watson optical potential for single scattering, which
is  of the form
\begin{equation}
\langle {\bf k}' | \langle \phi_A|PUP|\phi_A\rangle
{\bf k} \rangle\equiv U_{el}({\bf k}',{\bf k})=\sum_{i=N,P}\left
\langle {\bf k}' | \langle \phi_A|\hat\tau_{0i}({\cal{E}})|\phi_A\rangle
{\bf k}\right\rangle,
\label{eq:1}
\end{equation}
where $P$ is a projector on the ground state of the nucleus (we assume here $^6$He to
have 2 neutrons and protons in the s-shell, and 2 neutrons being in the $p_{3/2}$-shell),
and $\hat\tau_{0i}({\cal{E}})$ represents the nucleon-nucleon (NN) t-matrix evaluated at the
energy ${\cal{E}}$ of the system. The summation over $i$ indicates that one has
to sum over $N$ neutrons and $Z$ protons.  The structure of Eq.~({\ref{eq:1}}) 
is graphically shown in Fig.~1. The vectors {\bf k} and {\bf k'}
are the initial and final momenta of the projectile, while {\bf p} and {\bf p'}
are the initial and final momenta of the struck nucleon in the nucleus.
Changing variables to ${\bf q}={\bf k'}-{\bf k}={\bf p}-{\bf p'}$, ${\bf
K}=\frac{1}{2}({\bf k}'+{\bf k})$, and ${\bf P}=\frac{1}{2}({\bf p'}+{\bf p})$
as well as neglecting the recoil of the target nucleus leads to
\begin{equation}
U({\bf q},{\bf K})=\sum_{i=N,P} \int d^3{\bf P} \; {\hat \tau}_{0i} 
\left({\bf q}, \frac{1}{2}({\bf K}-{\bf P}); {\cal{E}}\right)\; 
\rho_i \left({\bf P} -\frac{{\bf q}}{2}, {\bf P} -\frac{{\bf q}}{2}\right).
\label{eq:2}
\end{equation}
The calculation of the first order optical potential relies on two basic
input quantities. One is the fully-off-shell NN t-matrix, which represents 
the current understanding of the nuclear force, and the other is the single
particle density matrix of the nucleus under consideration. 

The most general form of any NN t-matrix can be written as linear combination
of six spin-momentum operators~\cite{wolfenstein-ashkin,Fachruddin:2000wv} as
\begin{eqnarray}
\lefteqn{{\hat \tau}_{0i} ({\bf q},{\bf K}, {\cal E}) =} \cr
& & A ({\bf q},{\bf K}, {\cal E}) {\bf 1} +i C({\bf q},{\bf K}, {\cal E})
(\sigma^{(1)} + \sigma^{(2)}) \cdot {\hat {\bf n}}  \cr
&+&  M ({\bf q},{\bf K}, {\cal E}) (\sigma^{(1)} \cdot {\hat {\bf
n}})(\sigma^{(2)} \cdot {\hat {\bf n}})  + 
\left[ G ({\bf q},{\bf K}, {\cal E}) - H({\bf q},{\bf K}, {\cal E}) \right] 
(\sigma^{(1)} \cdot {\hat {\bf q}}) (\sigma^{(2)} \cdot {\hat {\bf q}}) \cr
&+& \left[ G ({\bf q},{\bf K}, {\cal E}) + H({\bf q},{\bf K}, {\cal E}) \right] 
(\sigma^{(1)} \cdot {\hat {\bf K}}) (\sigma^{(2)} \cdot {\hat {\bf K}}) \cr
&+& D ({\bf q},{\bf K}, {\cal E}) \left[ (\sigma^{(1)} \cdot {\hat {\bf q}})
(\sigma^{(2)} \cdot {\hat {\bf K}}) + (\sigma^{(1)} \cdot {\hat {\bf K}})
(\sigma^{(2)} \cdot {\hat {\bf q}}) \right],
\label{eq:3}
\end{eqnarray}
where the amplitude $D ({\bf q},{\bf K}, {\cal E})$ is zero on the energy shell.

\begin{figure}
\begin{center}
\includegraphics[width=70mm,angle=-90]{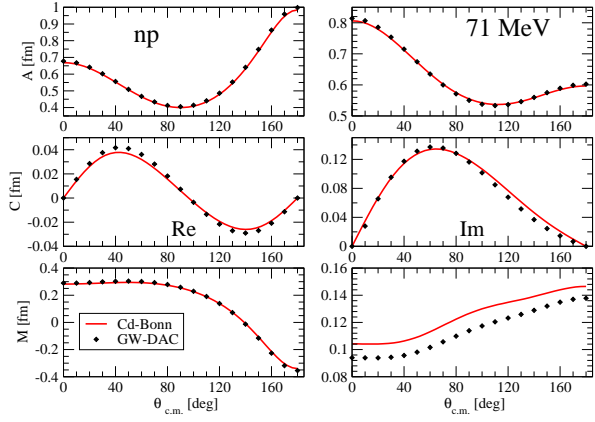}
\vspace{-2mm}
\caption{The Wolfenstein amplitudes for neutron-proton scattering at 71~MeV
laboratory NN kinetic energy based on the Cd-Bonn potential~\cite{Machleidt:2000ge},
which contribute to the optical potential for $^6$He.
The points show the extracted values from the GW-DAC
current analysis~\cite{GW-DAC,Arndt:2007qn}.
}
\end{center}
\label{fig2}      
\end{figure}
 
\begin{figure}
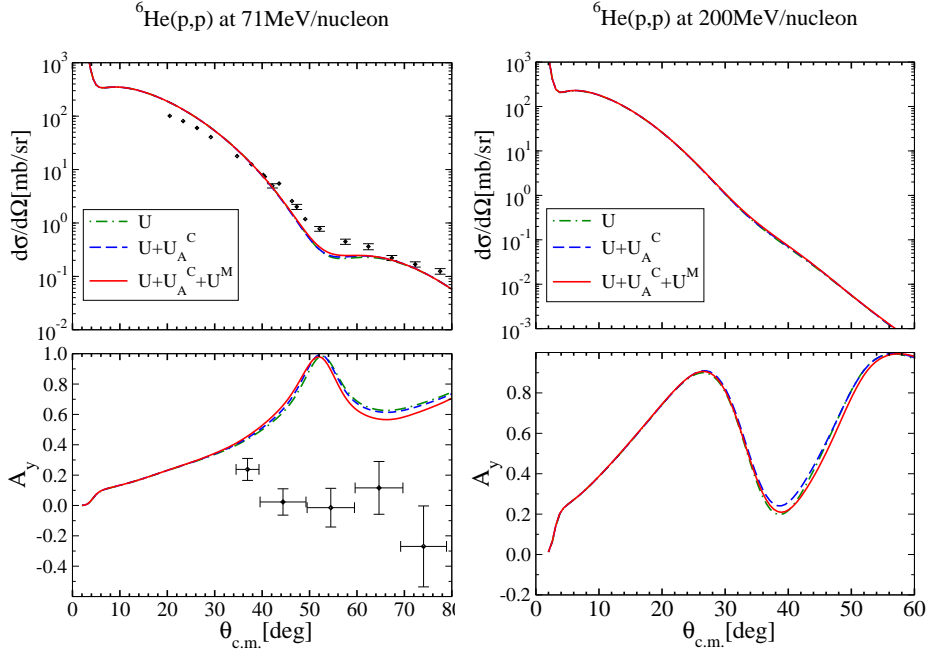

\includegraphics[width=60mm]{observ71CDB.eps}
\includegraphics[width=60mm]{observ200CDB.eps}
\vspace{1mm}
\caption{The angular distribution of the differential cross section (top panels) and the 
analyzing power ($A_y$) (bottom panels) for elastic scattering of $^6$He from protons at
projectile laboratory energies of 71~MeV/nucleon (left) and 200~MeV/nucleon (right). The
calculations show a first order optical potential based on the Cd-Bonn
potential~\cite{Machleidt:2000ge}. The dash-dotted line represents a calculation with a
traditional optical potential derived by using the NN amplitudes $A$ and $C$ only, while 
for the dashed line and solid lines the contributions of the valance neutrons to the
central and spin-orbit terms are successively added. All calculations do not include the
target recoil. The experimental data are taken from Ref.~\cite{Sakaguchi:2011rp}.
}
\label{fig3}       
\end{figure}

Since our goal is to test the effect of the two valance neutrons on the scattering observables
rather than to make a detailed comparison with data, we approximate the density matrix for
$^6$He by two harmonic oscillator terms, as is e.g. used for the COSMA
densities~\cite{Zhukov:1994zz}, namely a fully occupied $s$-shell and two neutrons in the
$p_{3/2}$-shell. The oscillator constants are $\nu_s=0.355$~fm$^{-2}$ corresponding to a
$^6$He charge radius of 2.054~fm~\cite{Wang:2004ze} and $\nu_p=0.3225$~fm$^{-2}$ corresponding 
to a matter radius of 2.32~fm~\cite{lbwang}. Explicitly, the radial wave functions are given as
\begin{eqnarray}
\psi_s^m (p) &=&  (2\pi)^{3/2} \left(\frac{4}{\sqrt{\pi} \nu_s^{3/2}}\right)^{1/2} \frac{1}{4\pi}
e^{-p^2/2\nu_s} {\cal Y}^{\frac{1}{2}}_{0m} ({\hat p}) \cr
\psi_p^m (p) &=& (2\pi)^{3/2} \left(\frac{4}{\sqrt{\pi} \nu_p^{3/2}}\right)^{1/2}
\sqrt{\frac{2}{3}} \frac{p}{\sqrt{\nu_p}} e^{-p^2/2\nu_p} \; {\cal Y}^{\frac{3}{2}}_{1m}({\hat p}),
\label{eq:4}
\end{eqnarray}
from which the s-shell and p-shell single particle densities are obtained. Specifically,
the p-shell projected on the $j=3/2$ state is given as
\begin{equation}
\psi_{p_{3/2}}(p) := f_{p_{3/2}}(p) \frac{1}{\sqrt{4}}\left( {\cal
Y}^{\frac{3}{2}}_{1\frac{3}{2}} ({\hat p}) - {\cal Y}^{\frac{3}{2}}_{1\frac{1}{2}}({\hat p}) 
+ {\cal Y}^{\frac{3}{2}}_{1-\frac{1}{2}}({\hat p}) - {\cal Y}^{\frac{3}{2}}_{1-\frac{3}{2}}({\hat p})\right),
\label{eq:5}
\end{equation}
where the spin-angular momentum functions are denoted by ${\cal Y}^j_{lm}$ and the radial
part by $f_{p_{3/2}}(p)$.
When calculating the optical potential of Eq.~(\ref{eq:1}), expectation values of the
spin-momentum operators $\sigma^{(2)} \cdot {\hat {\bf k}_i}$, with ${\bf k}_i \equiv {\hat
{\bf n}}, {\hat {\bf q}}, {\hat {\bf K}}$, must be taken with all nuclear single-particle wave
functions. If the expectation value is taken with wave functions describing a closed shell, the
sum over all nucleons in that shell adds to zero. Thus, for closed shell nuclei only the
terms corresponding to the Wolfenstein amplitudes $A$ and $C$ in Eq.~(\ref{eq:3}) contribute to
the folding optical potential. For $^{16}$O such a calculation is given in
Ref.~\cite{Elster:1989en}. For an open-shell nucleus as $^6$He, those expectation values 
contribute the optical potential. We find by explicit calculation 
\begin{eqnarray}
\psi_{p_{3/2}}(p') \; \sigma^{(2)}\cdot {\hat {\bf n}} \; \psi_{p_{3/2}}(p)
&=& -i \frac{2 p p'}{9\pi^{3/2}\nu_p^{5/2}} \; e^{-\frac{1}{2\nu_p}(p^2+p'^2)} \sin \alpha_{pp'}
\cos \beta \cr
\psi_{p_{3/2}}(p') \; \sigma^{(2)}\cdot {\hat {\bf K}} \; \psi_{p_{3/2}}(p) 
&=& -i \frac{2 p p'}{9\pi^{3/2}\nu_p^{5/2}} \; e^{-\frac{1}{2\nu_p}(p^2+p'^2)} \sin \alpha_{pp'}
\cos \delta \cr
\psi_{p_{3/2}}(p') \; \sigma^{(2)}\cdot {\hat {\bf q}} \; \psi_{p_{3/2}}(p) &=& 0,
\label{eq:6}
\end{eqnarray}
where $\cos \beta$ is the angle between the normal vectors in the nucleon-nucleus (NA) and
NN frame, $\alpha_{pp'}$ is the angle between {\bf p} and {\bf p'}, and $\cos \delta$ is the angle
between the total momentum in the NN frame and the normal in the NA frame.
This means that all amplitudes from the NN t-matrix, Eq.~(\ref{eq:3}), 
contribute to the optical potential when integrated with the single-particle 
density matrix of the $p_{3/2}$-shell neutrons. 
The term proportional to $i C({\bf q},{\bf K}, {\cal E}) \;
\sigma^{(2)}\cdot {\hat {\bf n}}$ has no dependence on the spin of the projectile, $\sigma^{(1)}$, 
and thus will give an additional contribution to the central part of the optical potential. 
All other terms contain a scalar product of $\sigma^{(1)}$ with a momentum. The operator
 for the amplitude $M({\bf q},{\bf K}, {\cal E})$ has the typical structure of a
spin-orbit operator and only contributes to the spin-orbit potential, while the terms
proportional to $(\sigma^{(1)}\cdot {\hat {\bf K}})$ and 
$(\sigma^{(1)}\cdot {\hat {\bf q}})$  contribute to the central as well as to the
spin-orbit potential.  The explicit integration over the single-particle wave functions
for the valence neutrons and the full NN t-matrix of Eq.~(\ref{eq:3}) reveals that 
the integrals involving the amplitudes $(G+H)$ and $D$ give a zero contribution to the
optical potential of the valence neutrons.  Therefore, the final expression for the
optical potential for 
$^6$He scattering off protons has the following structure
\begin{equation}
U_{^6He} ({\bf q},{\bf K}) = \sum_{i=N,P} U_{core} ({\bf q},{\bf K}) + U_{val} ({\bf q},{\bf K}),
\label{eq:7}
\end{equation} 
where $U_{core}$ is the usual optical potential for a closed shell nucleus. The optical
potential due to the valence neutrons acquires additional terms in its central as well as
its spin-orbit part,
\begin{eqnarray}
U_{val_{central}} &=& U_A({\bf q},{\bf K}) + U_A^C ({\bf q},{\bf K}) \cr
U_{val_{spin-orbit}} &=&  U_C ({\bf q},{\bf K}) + U^M ({\bf q},{\bf K}).
\label{eq:8}
\end{eqnarray}

\section{Discussion and Outlook}
\label{sec:2}

Our calculations of the optical potential for $^6$He scattering off protons are carried out with
the harmonic oscillator density given in Section~\ref{sec:1} and a NN t-matrix derived from the
Cd-Bonn potential~\cite{Machleidt:2000ge}. The Wolfenstein amplitudes for neutron-proton (np)
scattering at 71~MeV laboratory kinetic energy are shown in Fig.~\ref{fig2} as function of the
np c.m. angle. The amplitude $A$ gives the major contribution to the central term, while 
the imaginary part of $C$ is responsible for the real part of the spin-orbit term. 
In addition, the valence neutrons contribute through the $M$ amplitude to the spin-orbit term,
as well as to the central term through $C$.     
Since the expectation values of the spin-momentum operators for the neutron in the
$p_{3/2}$-shell of $^6$He contain a factor $\sin \alpha_{pp'}$, the forward direction of the
amplitudes does not contribute to the optical potential.   

The calculations of the angular distributions of the differential cross section and the analyzing
power $A_y$ at 71~MeV/nucleon and 200~MeV/nucleon are shown in Fig.~\ref{fig3}.  The dash-dotted
lines show a `traditional' calculation with an optical potential as it would be derived for a
closed shell nucleus~\cite{Elster:1989en}. 
The dashed line gives the contribution of the valence neutrons to the central part of the
optical potential, while for the solid line their contributions to the cental and the spin-orbit
term are added. All optical potentials are calculated neglecting the target recoil, similar to 
 the calculations in Ref.~\cite{Elster:1989en}.  Overall, the additional contributions of the
valence neutrons to the optical potential are small. However, they show energy dependence. At
71~MeV/nucleon, the contribution to the central term are negligible, only the contribution of
from the $M$ amplitude to the spin orbit term is visible. At 200~MeV/nucleon both contributions
are of the same size, but act in opposite directions so that the net contribution vanishes. 
This changes slightly when adding the effect of recoil. Our preliminary calculations show that
when including recoil, the effect on the spin-orbit term through the $M$ amplitude is slightly
enhanced, while the effect due to the additional central term slightly decreases. This will be
discussed in a future manuscript.

In previous work some of the authors investigated the optical potential for $^6$He derived in
a cluster description~\cite{Weppner:2011px}, in which the internal dynamics of $^6$He as alpha-core
plus two neutrons was taken into account explicitly. In this work it was found that the
description of the analyzing power at 71~MeV was sensitive to the cluster dynamics. However, it
was also found, that the quality of the optical potential for the alpha-core was important. 
In principle, the additional terms due to the open shell structure of the $p_{3/2}$-valence
neutrons would also have to be included for the two neutrons in the cluster description.
   
 Our present work mainly concentrates on taking into account effects related to the 
dynamics of the $^6$He nucleus as well as the interaction between the nucleons. In contrast to 
that our description of the single-particle density matrix is relatively simple. In a recent
work~\cite{Kaki:2012hr} a somewhat opposite approach was taken by deriving an optical potential
in the Glauber approximation, however employing highly sophisticated wave functions for the 
$^6$He nucleus.  It appears that a microscopic understanding of the reaction $^6$He(p,p)$^6$He
requires considering the dynamics of the reaction as well as the structure of $^6$He 
at an equal level of sophistication.

\end{document}